\documentclass[preprint]{aastex}

\def\ion#1#2{#1$\;${\small\rm{#2}}\relax}

\shorttitle{ }
\shortauthors{ }

\begin{document}


\title{The First Observations of the Second Brightest Quasar}


\author{K.\ M.\ Leighly}
\affil{Department of Physics and Astronomy, The University of
Oklahoma, 440 W. Brooks St., Norman, OK 73019; leighly@ou.edu}

\author{J.\ P.\ Halpern, D.\ J.\ Helfand}
\affil{Department of Astronomy, Columbia University, 550 W. 120th Street, 
New York, NY 10027 }

\author{R. H. Becker }
\affil{Physics Dept., University of California, Davis, CA 95616}
\affil{Institute of Geophysics and Planetary Physics, Lawrence
Livermore National Laboratory, PO Box 808, L--413, Livermore, CA
94551-9900}

\author{C. D. Impey}
\affil{Steward Observatory, University of Arizona, 933 North Cherry
Avenue, Tucson, AZ 85721}



\begin{abstract}
We report the discovery of a new bright quasar PHL 1811 as part of a
followup program to identify bright quasars in the {\it FIRST} radio
survey.  With B=13.9, R=13.9, and z=0.192, this quasar is now the
second brightest quasar (in apparent magnitude) known beyond z=0.1.
Optically classified as a Narrow-line Seyfert 1 galaxy (NLS1), PHL
1811 is unusual for an object in this class in that it was not
detected in X-rays in the {\it ROSAT} All Sky survey or in any
previous X-ray survey.  A follow-up {\it BeppoSAX} observation
confirms that it is deficient in X-rays compared with other quasars,
but poor signal to noise prevents a definitive characterization of the
X-ray spectrum.  We consider three alternative hypotheses to explain
why PHL 1811 is a weak X-ray source: 1.) it is a BALQSO and suffers
X-ray absorption; 2.)  like several other luminous NLS1s, it exhibits
high amplitude X-ray variability, and has been observed only when it
is in an X-ray quiescent state; 3.) it is intrinsically weak because
it simply lacks an X-ray emitting region.

\end{abstract}




\section{Introduction}
Bright, luminous quasars, relatively rare in the local universe, offer
a valuable laboratory for studying astrophysical phenomena.  Good
signal-to-noise observations can be obtained with a modest amount of
observing time, and these objects, which display the effects of
accretion onto supermassive black holes, can be studied in detail.  Bright
quasars can also be used to probe the interstellar and intergalactic
medium, as their brilliant continuum allows detection of faint
absorption lines.

For the past eight years, the NRAO Very Large Array (VLA) has been
used to construct Faint Images of the Radio Sky at Twenty centimeters.
The {\it FIRST} survey was designed to produce the centimetric equivalent of
the POSS-I Sky Survey, a uniform sky survey which surpassed existing
radio surveys by a factor of 50 in both sensitivity and angular
resolution. To date, nearly 8000 deg$^2$ of the North Galactic Cap
have been covered to a sensitivity limit of 1.0~mJy; 700,000 radio
sources have been detected, all of which have astrometric accuracies
of better than $1^{\prime\prime}$ at 90\% confidence (Becker et al.\
1995; White et al.\ 1997).

One of our major follow-up programs has been the {\it FIRST} Bright
Quasar Survey (FBQS: Gregg et al.\ 1996; White et al.\ 2000). Using
coincidences between {\it FIRST} radio sources and stellar objects on
the POSS-I plates brighter than 18th magnitude, spectra have been
obtained for $\sim 2000$ counterparts, leading to the discovery of
more than 1000 quasars and BL~Lac objects, over 80\% of which were
previously unknown.  Recently, extending this program to a strip in
the South Galactic Cap, we discovered an extremely
bright quasar at $z=0.192$.  Its magnitudes on the POSS I plates from
the USNO catalog (Monet et al.\ 1996) are $B = 13.9$ and $R = 13.9$,
and it is of similar brightness on a UK Schmidt plate taken in 1988.
Its absolute magnitude is $\rm M_V=-26.8$ (for $\rm H_0=50\,
km\,s^{-1}\,Mpc^{-1}$).  These properties make it the second brightest
quasar (in apparent magnitude) known with $z>0.1$ after 3C~273.  Its position is (J2000)
$21^{\rm h}55^{\rm m}01.\!^{\rm s}48,\
-09^{\circ}22^{\prime}24.\!^{\prime\prime}7$; a finding chart is given
in Figure 1.  Its radio flux density at 20 cm is only 1.4~mJy, so
although this object was first discovered in a radio survey, it is a
radio quiet quasar.  It is also a faint {\it IRAS} source.  The only
other mention of this object in the literature is as PHL~1811, one of
the many blue ``stars'' in the Palomar-Haro-Luyten plate survey (Haro
\& Luyten 1962).

In this paper, we report optical and X-ray observations of this new
bright quasar.

\section{Optical Spectroscopy}
PHL 1811 was observed at the Multiple Mirror Telescope on 1997
September 11 for 180 seconds.  The spectral resolution was about
1.95\AA\ per pixel.

The spectrum is shown in Figure 2.  It has been dereddened according
to the Cardelli, Clayton, \& Mathis (1989) law for Galactic reddening
of E(B$-$V)=0.046 magnitudes and shifted to the rest frame.  The
optical continuum is very blue.  While blends of lines make
characterization of the continuum difficult, we estimate that the
slope is $\alpha = -0.4 $ for $F(\nu)=\nu^{-\alpha}$ between 3050 \AA\
and 5600 \AA\/.  This is as blue as the bluest quasars observed (e.g.,
Grupe et al.\ 1998; also Grupe et al.\ 2000 \& Siebert et al. 1999)
and appears to be consistent with the continuum predicted by the
classical thin accretion disk model ($\alpha=-1/3$). The bluest part
of the spectrum may not be reliable as the airmass during the
observation was relatively high and the spectroscopic slit was not
rotated to the parallactic angle (Filippenko 1982).  However, this
effect would tend to lead to a loss of blue photons in general.

The optical spectrum shows the narrow H$\beta$, weak [\ion{O}{III}]
and strong \ion{Fe}{II} characteristic of a Narrow-line Seyfert 1
galaxy (Osterbrock \& Pogge 1985; Goodrich 1989).  The region around
H$\beta$ is cluttered with blends of \ion{Fe}{II}, making the analysis
of this region difficult, and therefore, to measure the width of the
H$\beta$ line, we perform an \ion{Fe}{II} subtraction using a template
obtained from the prototype NLS1 I~Zw~1. This procedure involves
shifting the high signal-to-noise template to the rest wavelength of
the quasar, broadening the template as necessary with a Gaussian, and
subtracting the amplitude-scaled template from the spectrum.  While
not perfect, this method has been frequently used to clear the region
around H$\beta$ of \ion{Fe}{II} lines (Boroson \& Green 1992; Grupe et
al.\ 1999; Leighly 1999b, among others).

The \ion{Fe}{II} subtracted spectrum is shown in the inset of Figure
1.  We fitted the H$\beta$ line with a Lorenzian profile and obtain a
FWHM of $1930 \,\rm km \, s^{-1}$.  The FWHM is less than $2000\,\rm
km\, s^{-1}$, indicating that quasar can be classified as a
Narrow-line quasar. H$\alpha$ is somewhat narrower, with FWHM of
$1650\,\rm km\, s^{-1}$.  The \ion{Fe}{II}-subtracted spectrum shows
that there is no detected [\ion{O}{III}] line.  The blue continuum and
lack of [\ion{O}{III}] have been observed in another luminous
Narrow-line Quasar, RX J1034.2$-$4258 (Grupe et al.\ 2000).  We
measure a \ion{Fe}{II} to H$\beta$ ratio of 1.1, making this object
moderately strong in \ion{Fe}{II} emission compared with other NLS1s
(e.g., Leighly 1999b).  \ion{He}{II}$\lambda 4686$ is also very weak
or undetectable.  We find that the ratio of H$\alpha$ to H$\beta$ is
2.66, a ratio consistent with that expected from case-B
recombination. This fact and the blue continuum suggest that there has
been no significant reddening of the spectrum.

We extrapolated our spectrum to 2500\AA\ in the rest frame and find
that the monochromatic flux is approximately $1.0 \times 10^{-25}\,\rm
erg\,cm^{-2}\,s^{-1}\,Hz^{-1}$.  For $H_0=50\rm \,
km\,s^{-1}\,Mpc^{-1}$, this corresponds to a luminosity of $2 \times
10^{31}\, \rm ergs\,s^{-1}\, Hz^{-1}$.  Wilkes et al.\ (1994) find an
average slope between 2500\AA\/ and $2\,\rm keV$ ($\alpha_{ox}$) of
1.62 for a quasar of this luminosity.

\section{X-ray Observations}

\subsection{{\it ROSAT}}

PHL 1811 is the second brightest quasar known beyond z=0.1.  Based on
the typical $\alpha_{ox}$ mentioned above, we expect that the
count rate in the {\it ROSAT} PSPC 0.15--2.4 keV bandpass to be $0.5
\, \rm cnts\, s^{-1}$ for a power law ($\Gamma=2.5$) plus Galactic
absorption ($N_H=3.65 \times 10^{20}\rm \, cm^{-2}$; Dickey \& Lockman
1990) model.  PHL 1811 should be a relatively bright quasar in X-rays,
and therefore it was a complete surprise to discover that this object
is undetected in the {\it ROSAT} All Sky Survey (RASS).

We use the RASS events files to estimate an upper limit on the
emission from this object. An examination of the events files
indicated that PHL~1811 was scanned over a period of only $\sim 34$
hours.  The exposure time listed for an object near PHL 1811 in the
{\it ROSAT} Bright Source Catalog is 334 seconds, and we use this
value for our estimate.  The PSF in the survey is degraded from the
nominal PSPC PSF at the aimpoint owing to the inclusion of data from
the edge of the detector where aberrations are significant.  We used a
bright nearby source to estimate the effective PSF, and found a
half-power radius of 111$^{\prime\prime}$.  After examining a large
source-free region, we estimated that there should be 3.6 background
photons in a region of this radius in 334 seconds.  We observe only
two photons in the region centered on the optical position of
PHL~1811.  Determining the upper limit on the source count rate
becomes a problem of joint probability distributions: what is the net
number of counts consistent with exactly two photons being observed
and a background rate of 3.6 at a particular confidence level?
Assuming the Poisson probability distribution applies, we find a 95\%
confidence upper limit of 2.2 photons.  The source flux then is 4.4
photons in 334 seconds, yielding an upper limit of $1.3 \times 10^{-2}
\rm \, ct\, s^{-1}$ in the 0.1--2.4 keV {\it ROSAT} PSPC bandpass.

\subsection{{\it BeppoSAX} Observation}

A 40~ks {\it BeppoSAX} observation was approved as part of the Cycle 4
program and the observation was performed on 2000 May 16.  We used the
default SAXDAS extracted events files for analysis.

The net exposure in the MECS detectors is 38 ks.  An adaptively
smoothed image is shown in Figure 3.  The position of the bright
central source is consistent with the optical position of the quasar.
We measure 183 photons between 1.65 and 10 keV in a $3^\prime$
extraction region centered on the quasar.  Following the method of
Fiore, Guainazzi \& Grandi (1999), we scale blank sky background to
match the background rate observed.  In the $3^\prime$ region, there
are 118 background photons, yielding 65 net counts in the MECS
spectrum.
 The object is only barely detected in the shorter LECS
observation (11.9 ks exposure).  We return to this point below.

Clearly the poor statistics make any detailed spectral analysis
unreliable.  We fitted the MECS spectrum with a power law and Galactic
absorption.  The resulting fit is acceptable ($\chi^2=6.5$ for 7
degrees of freedom).  We find a rather flat best-fitting photon index
of $\Gamma=1.5$; however, it is clearly not well constrained and the
$\Delta \chi^2 = 2.71$ errors of 0.9--2.3 overlap the canonical value
for AGN of $\sim 1.9$, as well as the very flat value near $\Gamma
\sim 1$ expected if the spectrum is dominated by Compton reflection
(e.g., Matt et al.\ 1996).  The observed 2--10 keV flux is $1.6 \times
10^{-13}\rm \, erg\, cm^{-2}\, s^{-1}$, and the intrinsic 2--10 keV
luminosity is $2.8 \times 10^{43} \,\rm erg\,
s^{-1}$. 

The power law model fits the spectrum adequately.  However, the ratio
of the data to the model shows a slight curvature that could indicate
absorption intrinsic to the quasar (Figure 4). Furthermore, we would
like to explore the range of intrinsic flux at 2 keV consistent with
the spectrum for the discussion of the spectral energy distribution in
Section 4. Adding absorption to the model, we find an improvement in
the fit ($\chi^2=3.3$ for 6 dof).  The best-fitting photon index is
3.3, and the column density is $10^{23}\rm \, cm^{-2}$ in the rest
frame of the quasar. These parameters are not well constrained.  When
we fix the photon index at $\Gamma=2.6$, approximately the steepest
value observed from NLS1s (e.g.\ Leighly 1999), and arguably the
steepest value plausible, then the absorption column is
$6.4^{+6.4}_{-3.9} \times 10^{22}\, \rm cm^{-2}$.  For these
parameters, the intrinsic 2--10 keV luminosity is $5.2 \times 10^{43}
\,\rm erg\, s^{-1}$.  Alternatively, if we require the unabsorbed flux
at 2~keV to match that required for the nominal value of
$\alpha_{ox}=1.6$ (see Section 4), a photon index of $4.4$ and an
absorption column of $1.6 \times 10^{23}\, \rm cm^{-2}$ are required,
and an acceptable $\chi^2$ of 3.2 for 7 degrees of freedom is obtained.

We used the LECS data to try to differentiate between these candidate
models. Using the {\it fakeit} command in XSPEC, we estimated the
number of photons expected in the 0.1--4.0 keV LECS bandpass in a
$4^\prime$ region.  For the power law plus Galactic absorption model,
16.5 photons are expected, and for the two models including intrinsic
absorption, 7--8 photons are expected.  We observe 45 photons, and
estimate that $36 \pm 3.1$ of them are background photons, by using
the average background scaling method discussed in Fiore, Guainazzi \&
Grandi 1999.  The number of source photons is then $9 \pm 7.4$,
showing that the LECS data cannot differentiate between models.

\section{The Spectral Energy Distribution of PHL 1811}

We constructed the spectral energy distribution (SED) in the rest
frame of PHL~1811 using our optical spectrum, the {\it ROSAT} upper
limits, the {\it BeppoSAX} results, as well as the {\it IRAS} flux
points and the 2MASS $J$, $H$ and $K_s$ filter photometry points
(Figure 5).  We do not plot the radio flux point; with $\nu,\nu
F(\nu)=(1.7 \times 10^{9}, 2.0 \times 10^{-17})$, the point is too far
off the scale of the plot.  We caution that the optical spectrum was
not made under photometric conditions; however, the optical flux is
consistent with that measured from the POSS plates and a UK Schmidt
plate and is therefore representative.  We note that the X-ray and
optical observations were not simultaneous. Therefore the derived SED
is not definitive but is probably representative.

The SED is shown in Figure 5.  A dotted line joins the extrapolated
flux at 2500\AA\ to the predicted flux at 2 keV computed using
$\alpha_{ox}=1.6$ expected for the average quasar, and the dashed
vertical line shows the approximate range of $\alpha_{ox}$ observed
for quasars of this optical luminosity (estimated by eye from Figure 9
of Wilkes et al.\ 1994).  The {\it ROSAT} upper limit is shown for
three values of the assumed photon index (2.0, 2.5 \& 3.0).  The {\it
BeppoSAX} results are shown by flat and steep power laws obtained for
fits with intrinsic absorption and $\Gamma=2.6$, and without intrinsic
absorption.  For the power law plus Galactic absorption model, we find
$\alpha_{ox}$=2.2.  The X-ray flux at 2~keV in the rest frame is a
factor of 34 below that expected for the average value of
$\alpha_{ox}$.  For the power law ($\Gamma=2.6$) plus intrinsic
absorption model, the inferred $\alpha_{ox}$ is 1.96, and the X-ray
flux is a factor of 8 below that expected for the average value of
$\alpha_{ox}$.  When we allow the absorption to increase to the
$\Delta\chi^2=2.71$ limit of $1.3 \times 10^{23}\rm\, cm^{-2}$, the
inferred X-ray flux is still at least 5 times smaller than  the nominal
value.  Therefore, the data strongly suggest that, even after possible
absorption is accounted for, PHL~1811 is intrinsically X-ray weak.

\section{Discussion}

PHL 1811 is classified  optically as a Narrow-line Seyfert 1
galaxy with a quasar luminosity.  Narrow-line Seyfert 1 galaxies are
generally bright soft X-ray sources, comprising up to 50\% of soft
X-ray-selected samples of AGN (e.g.\ Stevens 1989; Grupe et al.\
1999).  Based on average quasar properties, we expect PHL 1811 to be
34 times brighter at 2~keV than it was observed to be in our {\it
BeppoSAX} observation and than implied by the {\it ROSAT} All Sky
Survey upper limit.  We speculate on why it may be so faint in
X-rays below.

\subsection{Is PHL 1811 the Brightest BALQSO?}  

Broad-absorption line quasars are recognized by deep broad absorption
lines in their UV spectra.  It has been shown that BALQSOs are
typically either weak X-ray sources or not detected at all (e.g.\
Green \& Mathur 1996 for {\it ROSAT} pointed observations; Gallagher
et al.\ 1999 for {\it ASCA} observations).  More recently, it has been
demonstrated that many objects that appear to be deficient in soft
X-rays also show absorption lines in their UV spectra (Brandt, Laor \&
Wills 1999).  In fact, these authors find that the equivalent width of
the \ion{C}{IV}$\lambda$1549 absorption line is inversely correlated
with $\alpha_{ox}$, where in this case $\alpha_{ox}$ is defined as the
slope between 3000\AA\/ and 2~keV.  For our data, this alternative
$\alpha_{ox}$ is 2.15 and, based on Brandt, Laor \& Wills (1999)
results, we can expect PHL 1811 to be a BALQSO and have a \ion{C}{IV}
absorption line equivalent width $>20$\AA\/.  If this turns out to be
the case, PHL 1811 will be the brightest BALQSO known by far. PHL~1811
is probably not a low-ionization BALQSO, though, as there is no
evidence for absorption lines blueward of \ion{Mg}{II} in our optical
spectrum.

If PHL 1811 is a BALQSO, then the X-ray emission that we see may be from a
very steep spectrum that is very highly absorbed, although the
required photon index of 4.4 is so steep that it seems unphysical.
Alternatively, the intrinsic spectrum could be completely absorbed,
and the photons that we see have been scattered towards us by an
electron scattering mirror (e.g.\ Gallagher et al.\ 1999). We note
that the polarization properties of this quasar are not yet known.

\subsection{Is PHL 1811 the Brightest Variable Quasar-luminosity NLS1?}  

Besides being known for their typically bright soft X-ray emission,
Narrow-line Seyfert 1 galaxies are also known for their
high-amplitude, rapid variability.  Detailed analysis of NLS1 light
curves reveals, however, that the character of the variability is not
uniform across the class.  NLS1s with very strong soft excess
components in their {\it ASCA} spectra show very high amplitude X-ray
variability (Leighly 1999ab).  The light curves from these objects
consists of flares and quiescent periods, and during their quiescent
periods, the objects would be classified as X-ray weak.  Two objects
in particular appear to be deficient in X-rays even when their {\it
time-averaged} X-ray spectrum is considered: non-simultaneous {\it
HST} STIS and {\it ASCA} observations of IRAS~13224$-$3809 and
1H~0707$-$495 are consistent with a value of $\alpha_{ox}$ of 1.63
whereas the expected value is 1.4 for quasars of their
luminosity. There is little evidence for absorption in their X-ray
spectra (Leighly 1999b), and no evidence for intrinsic absorption
lines in their UV spectra (Leighly \& Halpern 2001). PHL~1811 is much
more luminous than these quasars, so it seems possible that it could
have been in the quiescent state for the entire {\it BeppoSAX}
observation or RASS scan period (elapsed time 24 and $\sim 34$ hours
respectively).

\subsection{Is PHL 1811 Intrinsically X-ray Weak?}

We have recently been examining several luminous NLS1s
identified in the {\it ROSAT} All Sky Survey and have found unusual
X-ray behavior that suggests they may become, intermittently,
intrinsically X-ray weak.  A particular example is RX~J2217.9$-$5941,
a luminous NLS1 similar to PHL~1811 in that it also has a very blue
optical spectrum and strong \ion{Fe}{II} emission (Grupe, Thomas \&
Leighly 2001). RX~J2217.9$-$5941 was bright during the {\it ROSAT} All Sky
Survey, but it faded through two subsequent {\it ROSAT} HRI
observations and an {\it ASCA} observation.  At the same time, the
optical flux remained constant, and the inferred $\alpha_{ox}$ at the
time of the last {\it ROSAT} HRI observation was 2.0, much steeper
than the expected value of 1.5 (Wilkes et al.\ 1994).  Although the
{\it ASCA} analysis was hampered by poor statistics and contamination
by nearby sources, we find no evidence for X-ray absorption.  Although
other explanations are possible, we suspect that RX~J2217.9$-$5941 is
now intrinsically X-ray faint.

We point out that there is no obvious reason why intrinsically X-ray
weak quasars should not exist. A QSO arguably cannot exist without
some kind of accretion disk as a source of fuel and also as a source
of ionizing photons to power the observed optical and UV emission
lines.  However, in a luminous quasar, the continuum emission observed
from the accretion disk itself is expected to be confined to the
optical and UV.  Hard X-ray emission, on the other hand, is thought to
be created in a corona that is separate from the accretion disk in 
luminous AGN.  When poorly-sampled light curves seemed to indicate
that X-ray and UV variability was correlated, it was thought that much
or perhaps all of the accretion energy was released in the corona, and
that the disk reprocessed some of that radiation into the optical and
UV bands (e.g.\ Svensson \& Zdziarski 1994).  However, recent
observations with better sampling show that hard X-ray reprocessing is
clearly not a dominant effect in some objects (Nandra et al.\ 1998).
Therefore, AGN do not need to have coronae, and they may be absent in
some objects.  The only reason to consider a corona necessary, in our
current understanding of AGN, is so that reprocessing can power the
tightly correlated UV and optical variability.  However, as there has
never been concentrated broad-band monitoring done on an object
similar to PHL 1811, so we cannot predict its broadband continuum
behavior.

Why should PHL 1811 lack a hard X-ray emitting region?  At this point,
we can only speculate.  As discussed by Laor (2000), narrow Balmer
lines in very luminous AGN may imply extremely high accretion rates,
if the width of the lines is dominated by virial motions, and
therefore PHL~1811 may have an extremely high accretion rate. We note
that RX~J2217.9$-$5941, which seems to have similar X-ray properties,
is also a very luminous NLS1. One possible geometry suggested for
accreting black holes is a central hot optically-thin
geometrically-thick X-ray emitting region, and a optically-thick
geometrically-thin optical and UV emitting accretion disk with large
inner radius.  At high accretion rates, it might be expected that the
inner radius of the accretion disk would stretch down to the innermost
stable orbit, and the central hot X-ray emitting region would not be
present.  Another possible geometry assumes an X-ray emitting corona
on top of the accretion disk, where the corona is fed by reconnecting
magnetic flux tubes that rise buoyantly from the accretion disk.  In
this case, perhaps the corona is present but is quenched by the strong
UV emission from the disk.  Alternatively, recalling that an accretion
disk is in some ways similar to a star, the feeding of the corona may
be disrupted, and the magnetic field strength may be less in a radiative
disk (similar to high mass stars) than in a convective disk (similar
to low mass stars).
  
UV spectroscopic observations will show us whether PHL~1811 has deep
absorption features characteristic of a BALQSO.  Such observations
have been approved as part of the {\it HST} Cycle 10 program, and {\it
FUSE} GTO observations are also planned.  Two approved {\it Chandra}
observations will help us understand whether or not the X-rays are
absorbed, and whether there is variability that would tend to
rule out scattering models for the X-rays.

\acknowledgments
The authors gratefully acknowledge the {\it BeppoSAX} team and in
particular Luigi Piro for significant help with this observation.  KML
acknowledges useful discussions with Andrzej Zdziarski. KML
gratefully acknowledges financial support by NASA-LTSA (NAG5-10171).
DJH \& RHR acknowledge support for the FIRST survey followup
observations from the NSF (AST 98-62732).  This research has made use
of the NASA/IPAC Extragalactic Database (NED) which is operated by the
Jet Propulsion Laboratory, California Institute of Technology, under
contract with the National Aeronautics and Space Administration. This
publication makes use of data products from the Two Micron All Sky
Survey, which is a joint project of the University of Massachusetts
and the Infrared Processing and Analysis Center/California Institute
of Technology, funded by the National Aeronautics and Space
Administration and the National Science Foundation.




\clearpage

\begin{figure}
\plotone{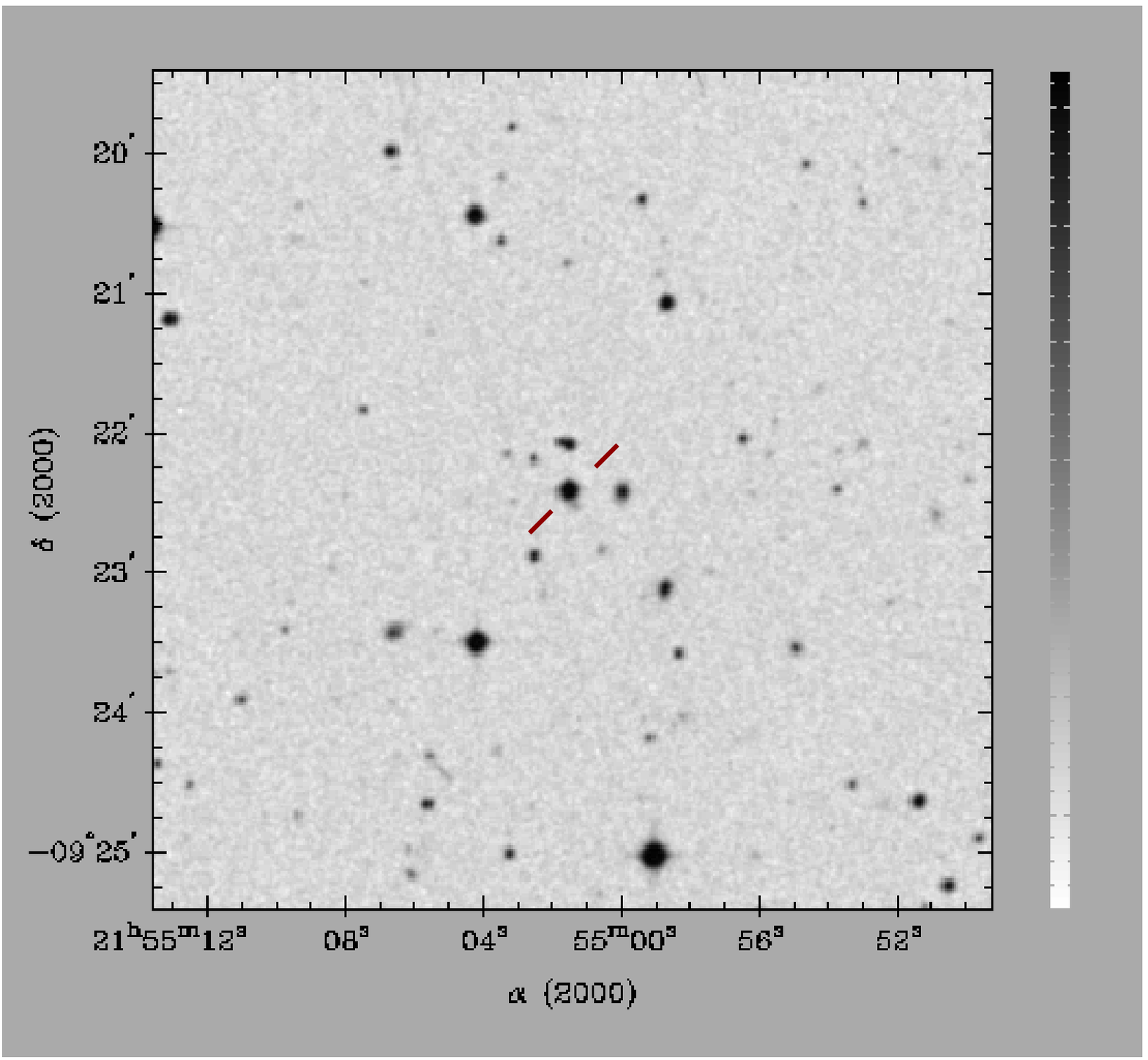}
\caption{A finding chart for PHL 1811 (J2000 $21^{\rm h}55^{\rm m}01.\!^{\rm
s}48,\ -09^{\circ}22^{\prime}24.\!^{\prime\prime}7$).}
\end{figure}

\begin{figure}
\plotone{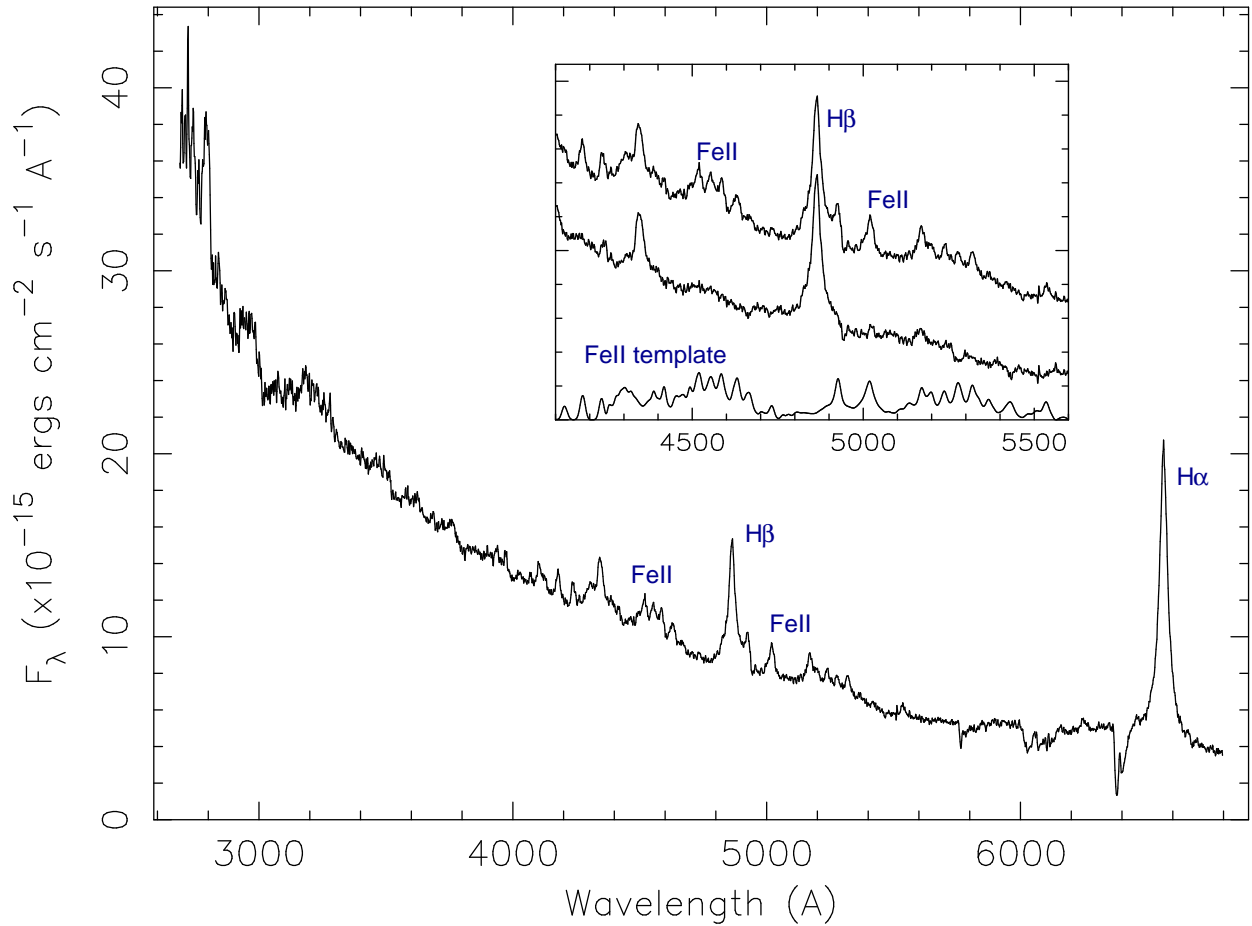}
\caption{The optical spectrum of PHL 1811.  The spectrum shows the
narrow Balmer lines, weak forbidden lines, and strong \ion{Fe}{II}
emission characteristic of a Narrow-line Seyfert 1 Galaxy. The
\ion{Fe}{II}-subtracted spectrum, shown as the middle frame in the
inset, reveals that the forbidden line [\ion{O}{III}]$\lambda 5007$,
typically one of the stronger lines in Seyfert spectra, is very weak
in PHL 1811.}
\end{figure}

\begin{figure}
\plotone{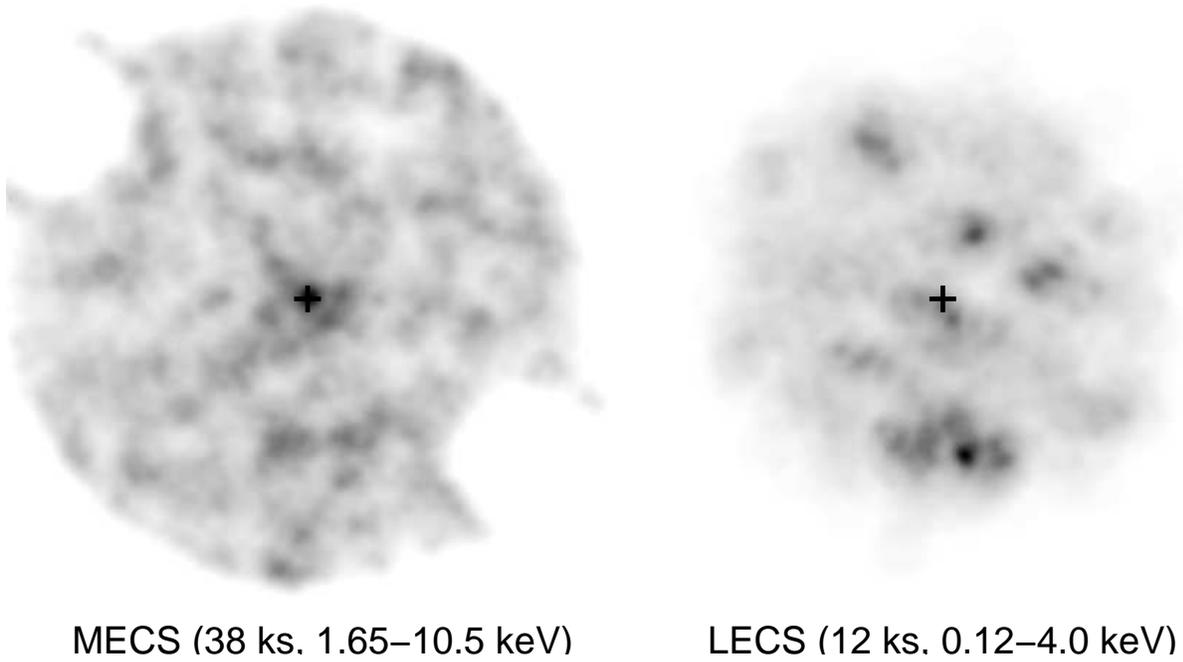}
\caption{Adaptively smoothed images from the {\it BeppoSAX}
observation of PHL 1811.  The image size is $55^\prime \times
55^\prime$.  The "+" marks the optical position of the quasar.  The
quasar is clearly detected in the MECS image, but only marginally
detected in the LECS image, probably because of the shorter exposure
and smaller effective area.}
\end{figure}

\begin{figure}
\plotone{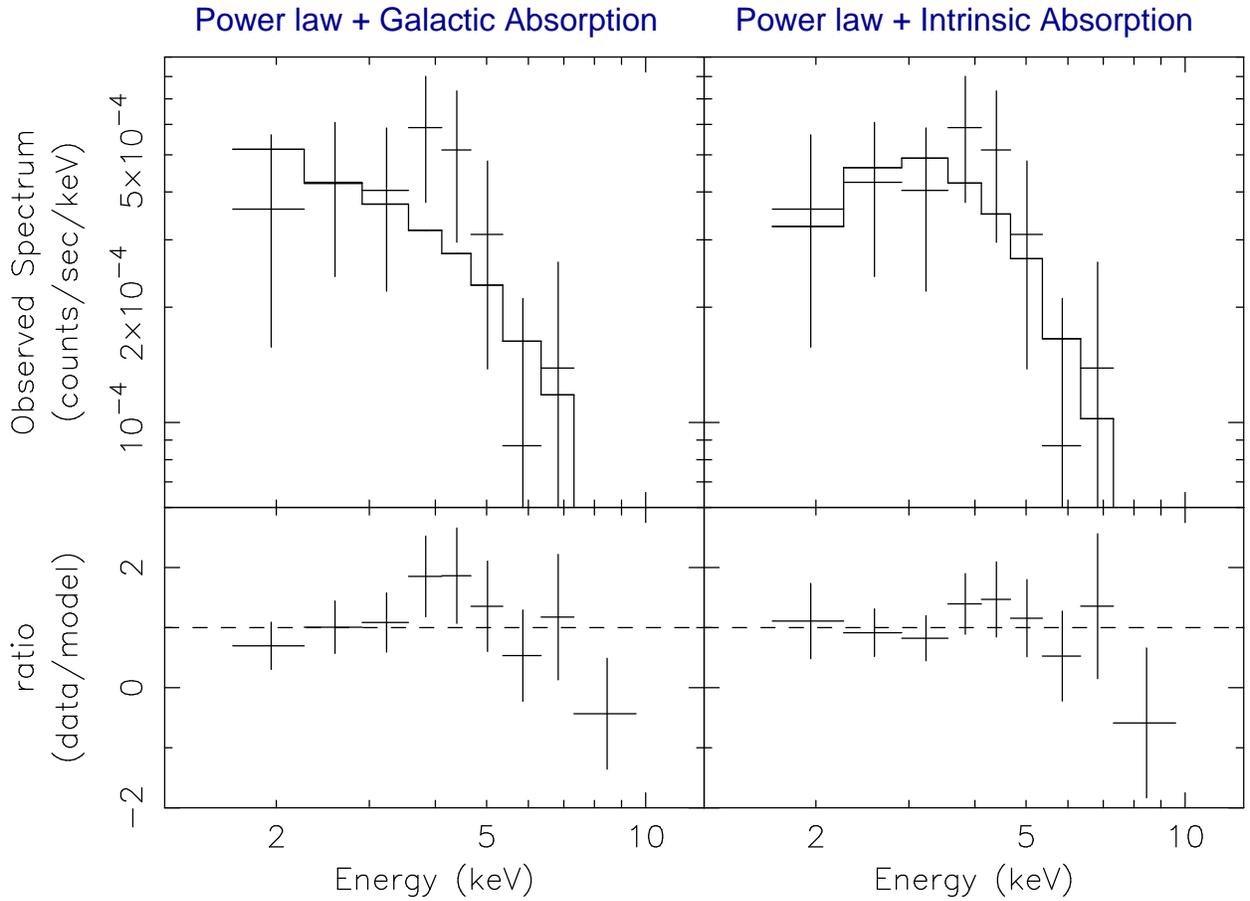}
\caption{Results of fitting the {\it BeppoSAX} MECS
spectrum.  The top panel shows the counts spectrum and the lower panel
shows the ratio of the data to model. The spectral parameters and
goodness-of-fit information are listed in the text.}
\end{figure}

\begin{figure}
\plotone{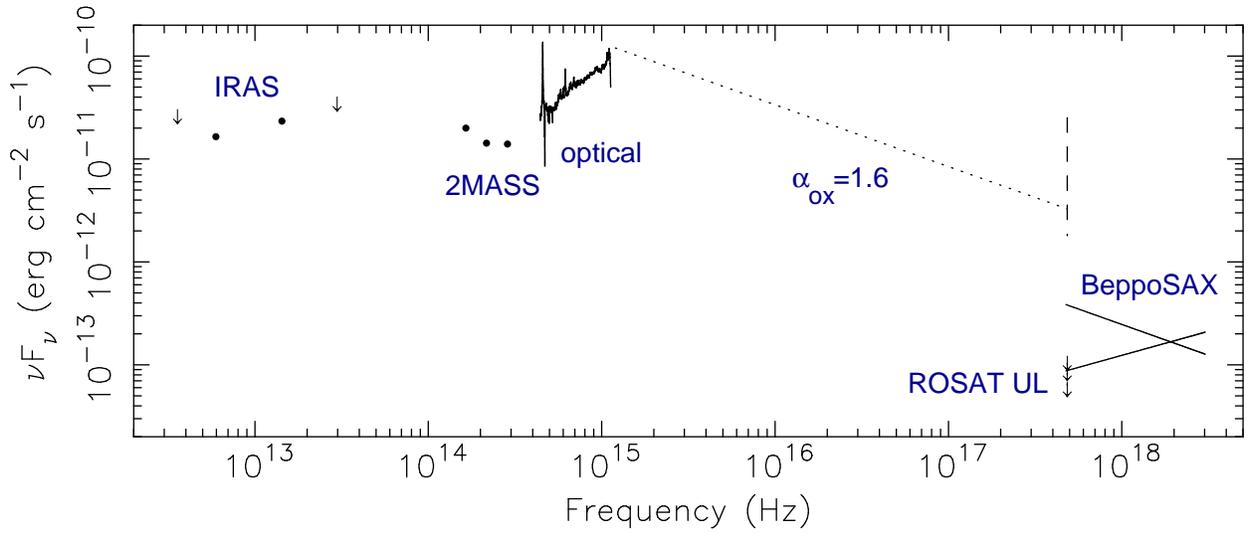}
\caption{The spectral energy distribution of PHL 1811, plotted as a
function of rest-frame frequency.  The dotted line denotes the 2 keV
flux for an average quasar; the observed flux is $\sim 34$ times
fainter than this.  The dashed line shows the range observed by Wilkes
et al.\ 1994. The {\it BeppoSAX} results are shown by flat and steep
power laws obtained for fits with intrinsic absorption and
$\Gamma=2.6$, and without intrinsic absorption. The {\it ROSAT} All
Sky Survey upper limits for a power law model with three assumed
photon indices are also shown.}
\end{figure}


\end{document}